\documentclass[a4paper,11pt]{article}
\usepackage{pos}
\usepackage{aas_macros}

\title{The light curve modeling of eclipsing cataclysmic variables: from period bouncers to nova-likes}
\ShortTitle{CVlab code V2.0}

\author*[a]{Sergey Zharikov}

\affiliation[a]{Instituto de Astronom\'ia, Universidad Nacional Aut\'onoma de M\'exico, \\ Apdo. Postal 877, Ensenada, Baja California, M\'exico, 22800}

\emailAdd{zhar@astro.unam.mx}

\abstract{I describe the basic concepts used in developing the computer code CVlab V2.0, which simulates the light curves of close binary systems with accretion-disk structures. I provide a detailed description of the code, including its underlying assumptions, limitations, and applications, as well as prospects for future development.
}

\FullConference{
THE GOLDEN AGE OF CATACLYSMIC VARIABLES AND RELATED OBJECTS - VII, September 1-6, 2025 in Palermo
(Italy)
}

\begin{document}
\maketitle

\section{Introduction}

Binary stars are among the most fundamental and informative objects in astrophysics. A significant fraction of all stars in the Galaxy—perhaps even the majority—reside in binary or multiple systems\citep{2013ARA&A..51..269D}. These systems provide one of the most powerful means of determining the fundamental stellar parameters that govern stellar structure and evolution. In particular, mass is the key quantity that dictates a star’s evolutionary path, internal structure, and final fate. For single stars, direct mass measurements are nearly impossible, whereas for binary systems, dynamical interactions allow the component masses to be determined with remarkable precision.
Observational studies of binary stars, combined with advances in light-curve modeling, spectroscopy, and astrometry, enable direct estimates of radii, effective temperatures, luminosities, and in some cases surface gravities and metallicities of the stellar components. Such measurements provide crucial tests for stellar evolutionary models. 
Close binary systems, in particular, play a central role in modern astrophysics. In these systems, the components are sufficiently close that tidal forces, irradiation, and mass transfer shape their observable characteristics and evolutionary trajectories. The exchange of matter and angular momentum produce a broad range of astrophysical phenomena, including accretion disks, hotspots, jets, nova eruptions, and in extreme cases, compact-object systems such as cataclysmic variables\citep{1995cvs..book.....W}, X-ray binaries\citep{2019NewAR..8601546K, 2023hxga.book..120B}, and millisecond pulsars\citep{2008LRR....11....8L, 2017JApA...38...42M}. Studying these systems offers insight not only into binary evolution, but also into the physics of accretion, magnetic fields, thermonuclear processes, and surface inhomogeneities in interacting stars.
 The presence of eclipses imposes strong geometric constraints on the inclination of the orbit, the separation of the components, and the distribution of radiation over the stellar surfaces. These constraints make eclipsing binaries some of the most accurate stellar laboratories known. When combined with radial-velocity measurements or distance estimates from Gaia\citep{2018AJ....156...58B}, eclipsing binaries can yield physical parameters with unprecedented precision, enabling stringent tests of fundamental stellar physics.
As the volume and quality of observational data continue to increase—with contributions from wide-field surveys such as ZTF\citep{2019PASP..131f8003B}, ASAS-SN\citep{2018MNRAS.477.3145J}, TESS\citep{2015JATIS...1a4003R}, and Gaia\citep{2016A&A...595A...1G}—the need for robust and flexible modeling tools becomes increasingly important. Modern modeling methods, built upon current understanding of binary-star physics, allow large samples of systems to be analyzed in a uniform way, providing statistically meaningful comparisons between theory and observations.
The development of specialized computational tools\citep{1971ApJ...166..605W, 2010MNRAS.402.1824C, 2012ApJ...748..115B, 2013A&A...549A..77Z, 2016ApJS..227...29P} is thus essential for interpreting the rich observational phenomenology of interacting binaries.

In this paper, I present the key characteristics of our modeling code, designed to derive the fundamental parameters of close binary systems containing accretion structures through detailed fitting of their observational light curves.

\begin{figure}
\includegraphics[width=0.5\textwidth, clip=]{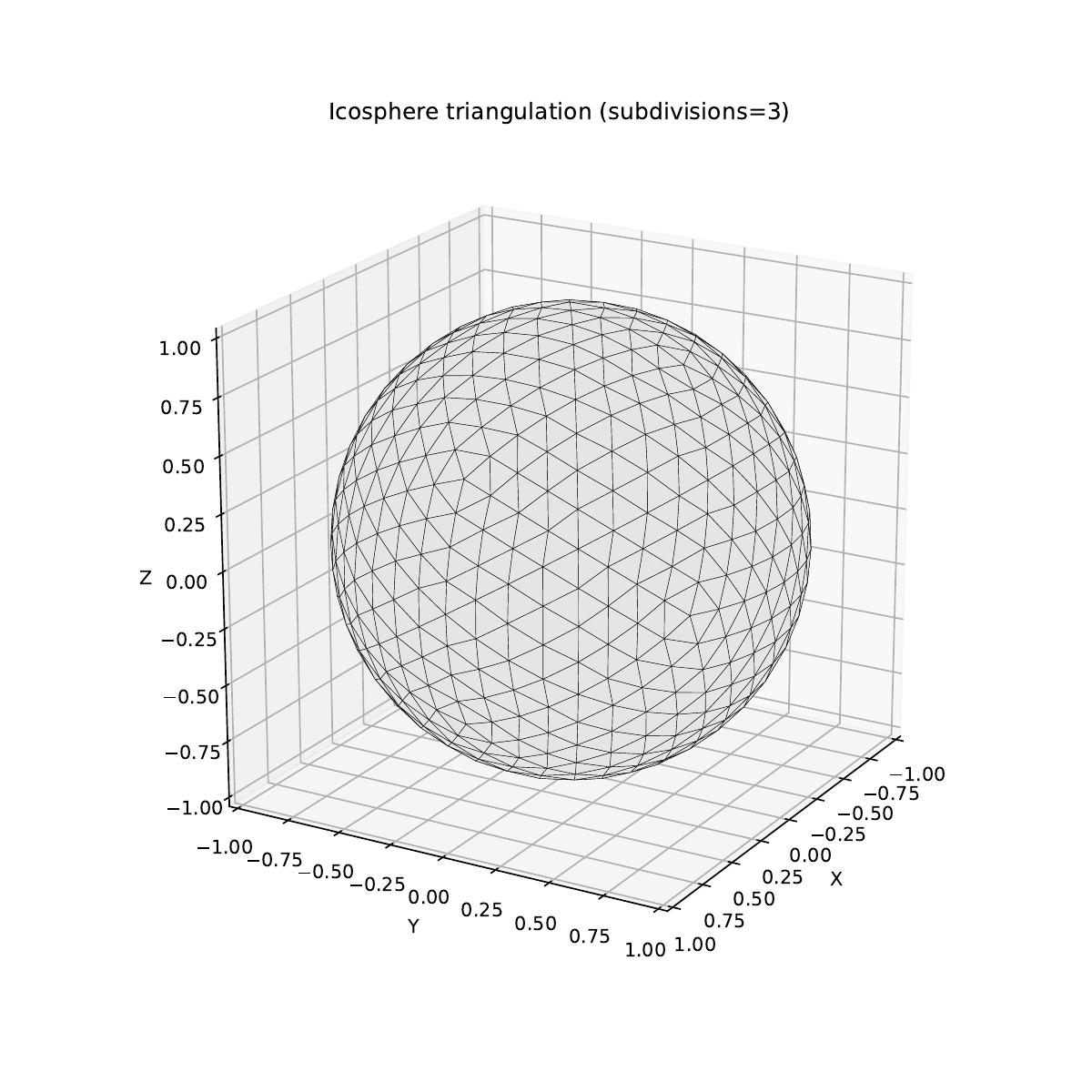}
\includegraphics[width=0.5\textwidth, clip=]{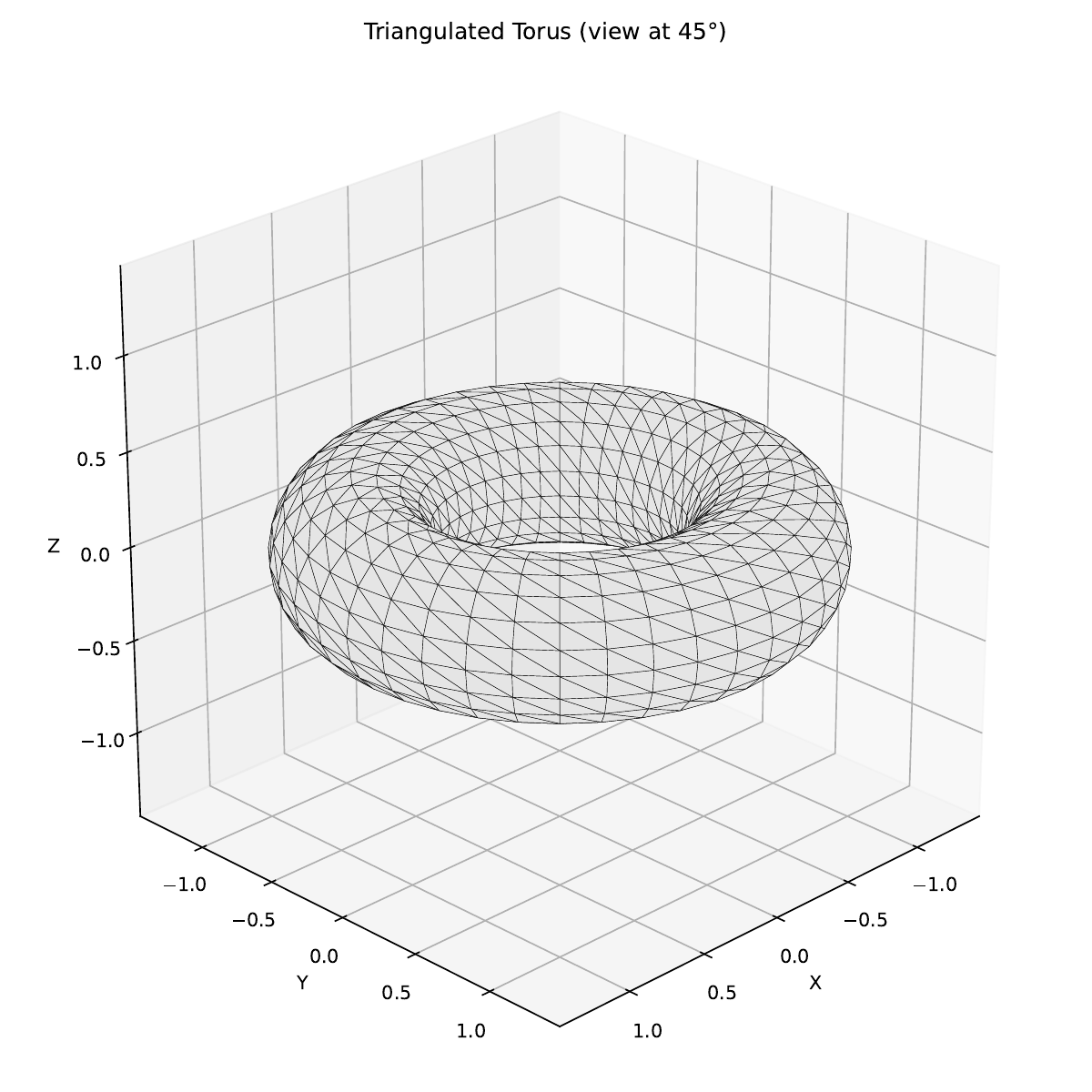}
\includegraphics[width=1.0\textwidth, bb = 1000 0 5000 3100, clip=]{view01S0003.pdf}
  \caption{Top)  The basic topological components of the code, the sphere and the torus, covered by non-overlapping triangular mesh. Bottom) Illustrative view on a binary system with different structure elements included in the code.}
  \label{fig1}
\end{figure}

\begin{figure}
\includegraphics[width=1.0\textwidth, bb = 100 300 600 525, clip=]{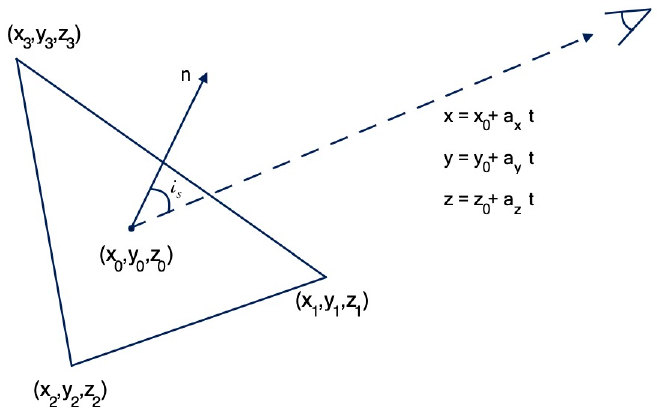}
  \caption{The illustration of basic triangle parameters.}
  \label{fig2}
\end{figure}

\section{Basic geometry assumption}
The emitting {\it surface} of an astrophysical source is defined as the plasma layer corresponding to an optical depth of $\tau = 2/3$ at a given wavelength or band.
In a general geometrical framework, the surface of any astrophysical object can be represented as a geometric transformation or combination of a small number of basic topological components—primarily a sphere and a torus, or their fragments (see Fig.~\ref{fig1}).
A non-overlapping triangular mesh of sufficient resolution can be used to approximate either of these surfaces with the required level of detail.
The emission from a point on such a surface toward the observer is described by the equation of a ray in three-dimensional space.
This ray is defined by the coordinates of the emitting point $(x_0, y_0, z_0)$ and by the components of the direction vector $(a_x, a_y, a_z)$ that is parallel to the line of sight:
\begin{equation}
 x = x_0+a_x*t,  y = y_0+a_y*t,  z = z_0+a_z*t,
\end{equation}
where $t$ is a scalar parameter representing the distance along the ray. 
Each \textit{surface element} from which the radiation originates is represented by a \textit{triangle} (see Fig.~\ref{fig2}) defined by its vertex coordinates 
$(x_1, y_1, z_1)$, $(x_2, y_2, z_2)$, and $(x_3, y_3, z_3)$ in three-dimensional space. 
The \textit{emitting point} associated with the triangle is taken to be its centroid, whose coordinates are given by
\begin{equation}
x_0 = \frac{1}{3}\sum_{i=1}^{3} x_i, \qquad
y_0 = \frac{1}{3}\sum_{i=1}^{3} y_i, \qquad
z_0 = \frac{1}{3}\sum_{i=1}^{3} z_i.
\end{equation}
For each triangular element, it is straightforward to determine the equation of its plane, 
its spatial orientation relative to the observer’s line of sight, and its surface area. 

 According to the principles of analytic geometry, each triangular element lies in a plane 
defined by the equation
\begin{equation}
A x + B y + C z + D = 0,
\label{planeEq}
\end{equation}
where $A$, $B$, and $C$ are the components of the normal vector ($\cos \alpha_x,~\cos\alpha_y,~ \cos\alpha_z$) to the plane, and $D$ 
is the constant term determined by the position of the triangle in space defined by
\begin{equation}
 A= cos (\alpha_x) = \begin{vmatrix}
y_1 & z_1 & 1 \\
y_2 & z_2 & 1 \\
y_3 & z_3 & 1 \\
\end{vmatrix},~ 
B=  cos (\alpha_y) = \begin{vmatrix}
z_1 & x_1 & 1 \\
z_2 & x_2 & 1 \\
z_3 & x_3 & 1 \\
\end{vmatrix},~
\end{equation}
\begin{equation}
C= cos (\alpha_z) =\begin{vmatrix}
x_1 & y_1 & 1 \\
x_2 & y_2 & 1 \\
x_3 & y_3 & 1 \\
\end{vmatrix},~~~~~~~~~~~~~~~
D = \begin{vmatrix}
x_1 & y_1 & z_1 \\
x_2 & y_2 & z_2 \\
x_3 & y_3 & z_3 \\
\end{vmatrix}
\end{equation}

The area of the triangular element can be expressed as
\begin{equation}
\Delta S = \frac{1}{2} 
\sqrt{(A^2 + B^2 + C^2)},
\end{equation}
where $A$, $B$, and $C$ are the coefficients of the plane equation 
(see Eq.~\ref{planeEq}) corresponding to the components of the surface normal vector.

The angle $\gamma_s$ between the line of sight and its projection onto the triangle surface 
is defined by
\begin{equation}
\sin \gamma_s = 
\frac{A a_x + B a_y + C a_z}
{\sqrt{A^2 + B^2 + C^2} \, \sqrt{a_x^2 + a_y^2 + a_z^2}},
\end{equation}
and the corresponding angle between the surface normal and the direction of the line is 
$i_s = \frac{\pi}{2} - \gamma_s$.

A given triangular element $j$ of a component is considered \textit{visible} to the observer 
if the ray emerging from its centroid $(x_0, y_0, z_0)$ does not intersect any internal point 
of another triangular element $k$ belonging to any other component of the system.

\section{Basic physic assumption}

\subsection{System geometry}
The geometry of a close binary system is determined by the shape of the Roche lobes. 
In a coordinate system corotating with the binary, with the origin at the primary component, 
the $x$-axis directed along the line connecting the mass centres of the two stars, 
and the $z$-axis perpendicular to the orbital plane, the Roche potential is given by 
the sum of the gravitational potentials of both stars and the effective potential 
arising from the fictitious centrifugal force\citep[and references therein]{1995cvs..book.....W}:

\begin{equation}
\Phi = - \frac{G M_1}{\sqrt{x^2+y^2+z^2}} - \frac{GM_2}{\sqrt{(x-a)^2+(y-a)^2+(z-a)^2}}-\frac{1}{2}\Omega^2_{orb}[(x-\mu a)^2+y^2],
\end{equation}
where $\mu = M_2/(M_1+M_2)$ and $\Omega_{orb} = 2\pi/P_{orb}$, $a$ is the system separation,  and $P_{orb}$ is the orbital period of the system $P_{orb}^2 = \frac{4\pi^2 a^3}{G(M_1+M_2)}$.
The inner Lagrangian point $L_1$ located between system component and is defined by $\partial\Phi/\partial x = 0$.
The mass ratio of the system is defined as:
\begin{equation}
q = \frac{M_2}{M_1}
    \end{equation}

\subsection{Element emission}
The current version of the code assumes that each surface element of the object emits 
as a blackbody with intensity $B(T_{\mathrm{eff}}^i, \lambda)$. 
The corresponding model flux received by the observer is

\begin{equation}
F_{\mathrm{model}}^k = \frac{1}{4\pi D^2} \sum_{i=1}^{N} F^i \times vis^i,
\label{fluxEq1}
\end{equation}
where
\begin{equation}
F^i = \pi \, \Delta S^i \cos(i_s^i)
\int_{\lambda_0}^{\lambda_1} f^{\mathrm{tr}}(\lambda)
B(T_{\mathrm{eff}}^i, \lambda) \, d\lambda.
\label{fluxEq2}
\end{equation}

Here $T_{\mathrm{eff}}^i$ is the effective temperature of the $i$-th surface element, 
$f^{\mathrm{tr}}(\lambda)$ is the filter transmission function obtained from the 
Filter Profile Service\footnote{\url{https://svo2.cab.inta-csic.es/svo/theory/fps3/index.php}}, 
and $vis^i = 1$ if the element is visible to the observer and $vis^i = 0$ otherwise. 
The index $k$ denotes the component of the system (e.g. primary, secondary, or accretion disc) 
to which the $i$-th surface element belongs.

The resulting synthetic magnitude in a photometric band $j$ is then given by
\begin{equation}
m_j = -2.5 \log_{10} \!\left(\sum_k F^k_j\right) - M_j + A_j[E(g-r)],
\end{equation}
where $F^k_j$ is the total flux from all visible elements of component $k$ in band $j$, 
$M_j$ is the photometric zero point for that band, and $A_j$ represents the interstellar 
extinction correction.

\subsection{The primary}
In the current version of the code, five types of primary stars are implemented: 
a zero-age main sequence (ZAMS) star, an evolved star, a white dwarf (WD), a neutron star (NS), 
and a black hole (BH). 
The shape of the primary component is always assumed to be spherical, 
which implies that  the Roche-lobe radius of the primary 
$R_{\mathrm{R.L.},1} \gg R_1$. 
This assumption is valid for compact objects and usually holds for ZAMS stars, 
but it should be reconsidered in systems where the primary is an evolved star.

For a ZAMS star, the stellar parameters follow the approximate relations 
$R_1 \propto M_1$ and $T_{\mathrm{eff},1} \propto M_1$, 
as given by \citep{2007MNRAS.382.1073M}. 
In the case of an evolved star, both the radius $R_1$ and the mass $M_1$ 
are treated as free parameters and $R_1 > R_1^{ZAMS}$. 

For a white dwarf primary, the effective temperature $T_{\mathrm{eff,WD}}$ 
is a free parameter, while its radius $R_{\mathrm{WD}}$ (in cm) 
is related to the mass $M_{\mathrm{WD}}$ (in solar masses) 
according to the mass--radius relation from \citep{1972ApJ...175..417N}:
\begin{equation}
R_{\mathrm{WD}} = 0.779 \times 10^9 
\sqrt{
\left( \frac{M_{\mathrm{WD}}}{1.44\,M_\odot} \right)^{-2/3}
-
\left( \frac{M_{\mathrm{WD}}}{1.44\,M_\odot} \right)^{2/3}
} (cm).
\end{equation}

The neutron star (NS) is assumed to have a fixed radius of 13~km, 
while its mass and effective temperature can be treated as free parameters.

The black hole (BH) is defined by its Schwarzschild radius:
\begin{equation}
R_{\mathrm{BH}} = \frac{2GM_{\mathrm{BH}}}{c^2}.
\end{equation}

For ZAMS or evolved stars, as well as for WDs and NSs, 
the primary component can possess a dipolar magnetic field 
and a surface spot, which are described in the following sections.

The rotational period of the primary (ZAMS, evolved star, or WD) is a free parameter, defined as a fraction of the orbital period.

\subsubsection{Limb and gravitational darkening}

The code includes the quadratic limb-darkening law for the primary and secondary,
\begin{equation}
   \frac{I_{\tilde{\mu}}}{I(1)} = 1 - \tilde{a}\,(1 - \tilde{\mu}) - \tilde{b}\,(1 - \tilde{\mu})^2, 
   \qquad \tilde{\mu} = \cos(i_s),
\end{equation}
with the limb-darkening coefficients $(\tilde{a},\,\tilde{b})$ taken from 
\citep{2012A&A...546A..14C, 2013A&A...552A..16C} for ZAMS and evolved stars.
Gravity darkening is implemented following \citep[equation~5.4]{2018maeb.book.....P}. 
In the case of a white dwarf  primary, the limb-darkening coefficients were taken from \citep{2020A&A...634A..93C}.

The corresponding coefficients are recalculated for the exact effective temperature, 
surface gravity ($\log g$), and photometric filter using linear interpolation between 
the nearest tabulated values.

\subsection{The secondary}

The shape of the stellar surface is determined by the gravitational potential within the corresponding Roche lobe. 
The maximum possible stellar radius corresponds to the case when the star completely fills its Roche lobe. 
Mass transfer between the components occurs when the secondary star fills its Roche lobe. 
If the secondary does not fill its Roche lobe, its surface shape follows the equipotential surface 
corresponding to the adopted filling factor. 
The filling factor is defined as the ratio between the distance from the center of mass of the secondary 
to the inner Lagrangian point $L_1$ and the distance corresponding to the full Roche-lobe radius.

The effective temperature of the secondary, $T_{\mathrm{eff,2}}$, is treated as a free parameter in the code. The relation\footnote{\url{http://www.pas.rochester.edu/~emamajek/EEM_dwarf_UBVIJHK_colors_Teff.txt}} for Spectral Type of the  secondary vs. $T_{\mathrm{eff,2}}$ follows to the definition of \citep{2013ApJS..208....9P}. 
It is assumed that the rotational period of the secondary is synchronized with the orbital period of the system.

\subsubsection{Irradiation of the secondary}

1) Diskless systems. \\

The irradiation of the secondary was taken into account as
\begin{equation}
T_2^d = T_2^n \left( 1 + \frac{F_{in}}{\Delta S \sigma (T_2^n)^4}\right)^\frac{1}{4}
\end{equation}
where $T_2^d$ and $T_2^n$ are effectuve temperatures of the "day-side" and "night-side" surface element of the secondary, respectively, $F_{in}$ is the effective heating flux incoming to the "day-side" surface element area $\Delta S$ of the secondary, and $\sigma$ is the Stefan-Boltzmann constant. 
The effective heating in such model defined by the irradiation factor $K_{irr}$ [ergs s$^{-1}$ cm$^{-2}$ sr$^{-1}$] as 
\begin{equation}
F_{in} = \cos(\alpha_{norm}) \Omega \Delta S K_{irr}
\end{equation}
where $\alpha_{norm}$ is the angle between the incoming wind/flux and the normal to the surface element $\Delta S$, $\Omega = \pi R_1^2/a^2$ is the solid angle from which the primary is visible from the surface element of the secondary.\\

2) Disk systems.\\ 
In the case of a disk system, not all parts of the primary surface are necessarily visible from a surface element on the irradiated (“day-side”) hemisphere of the secondary. For an optically thick accretion disk, the disk can cast a shadow on the secondary star’s surface.  
In the current version of the code, it is assumed that the variation of the irradiation factor $K_{\mathrm{irr}}$ also accounts for the partial visibility of the primary surface as seen from the secondary. The shadowed region is determined by the geometry of the disk surface (defined as the layer with optical depth $\tau = 2/3$).

\begin{equation}
F_{\mathrm{in}} = \frac{1}{2} \cos(\alpha_{\mathrm{norm}})\, \Omega\, \Delta S\, K_{\mathrm{irr}},
\end{equation}

\subsection{The accretion stream}
The equations of motion for a test particle originating from the secondary and moving in the $x$–$y$ plane of the Roche lobe of the primary are given by \citep{1976IAUS...73..237L} as
\begin{eqnarray}
    \frac{d^2 x}{dt^2} & = & 2\,\frac{dy}{dt} + x + 1 - \mu + \frac{(1 + x)(\mu - 1)}{r_2^3} - \frac{\mu x}{r_1^3}, \nonumber \\
    \frac{d^2 y}{dt^2} & = & -2\,\frac{dx}{dt} + y \left( 1 + \frac{\mu - 1}{r_2^3} - \frac{\mu}{r_1^3} \right),
\end{eqnarray}
where $r_2^2 = (x + 1)^2 + y^2$ and $r_1^2 = x^2 + y^2$. The coordinate system is centered on the primary (mass $M_1$) at the origin $(0, 0, 0)$, while the secondary (mass $M_2$) is located at $(-1, 0, 0)$.

In the model, the stream is assumed to have a constant cross-sectional size (a free parameter) along its trajectory. Its temperature varies linearly along the stream, starting from the value at the inner Lagrangian point ($L_1$), equal to the effective temperature of the secondary ($T_{\mathrm{eff,2}}$), and reaching $T_{\mathrm{stream}} = T_{\mathrm{hot\,spot}}$ at the disk–stream impact region (the “hot spot” or “hot line”).  
The emitted flux from the stream is defined by Equation~\ref{fluxEq2}.

\subsection{The accretion disk}

The accretion disk in the model is assumed to be circular and is characterized by its inner and outer radii, $R_{\mathrm{d,in}}$ and $R_{\mathrm{d,out}}$, respectively, and by the disk height $z_{\mathrm{d}}(r_{\mathrm{out}})$ at the outer edge.  
The vertical structure of the disk is described by
\begin{equation}
z_{\mathrm{d}}(r) = z_{\mathrm{d}}(r_{\mathrm{out}}) 
\left( \frac{r}{r_{\mathrm{out}}} \right)^{\gamma_{\mathrm{disk}}},
\end{equation}
where $\gamma_{\mathrm{disk}}$ is a free parameter. As an initial value, we adopt the standard value $\gamma_{\mathrm{disk}} = 9/8$ 
\citep[Eq2.51b]{1995cvs..book.....W}.

The model assumes that the disk radiates as a blackbody at the local effective temperature, which varies radially according to
\begin{eqnarray}
T_{\mathrm{eff}}(r) &=& T_0 
\left\{
\left(\frac{r}{R_1}\right)^{-3}
\left[ 1 - \left( \frac{R_1}{r} \right)^{1/2} \right]
\right\}^{\mathrm{EXP}}, \nonumber \\
T_0 &=& 
\left[
\frac{3 G M_1 \dot{M}}{8 \pi \sigma R_1^3}
\right]^{1/4},
\end{eqnarray}
where $R_1$ is the radius of the primary, $\dot{M}$ is the mass-transfer rate, and $\sigma$ is the Stefan–Boltzmann constant.  
In the standard steady-state accretion disk model, the exponent $\mathrm{EXP}$ is fixed at $0.25$ 
\citep[Eq.~2.35]{1995cvs..book.....W}; however,  we allow it to deviate slightly from this canonical value, following the approach of \citep{2010ApJ...719..271L}.  
The local flux from each disk surface element is then computed using Equation~\ref{fluxEq2}.

The equality of the inner and outer radii  $R_{\mathrm{d,in}} = R_{\mathrm{d,out}}= R_{B}$ of the disk defines a diskless system, in which, in the presence of a magnetic field at the primary (polar), this radius determines the radius of the magnetosphere $R_{B}$. At this distance, the transition occurs from the ballistic trajectory of the accretion stream to the motion of plasma along the magnetic field lines toward the magnetic pole of the primary.

\subsubsection{Disk limb-darkening}
We used linear limb-darkening law
\begin{eqnarray*}
\frac{I(\tilde\mu)}{I(1)} =  1 - u(1 - \tilde\mu) 
\end{eqnarray*}
where $u$ was taken $u=0.6$ following Eddington approximation (see \citep{1980AcA....30..127P}).

\begin{figure}
\includegraphics[width=1.0\textwidth, bb = 100 10 600 525, clip=]{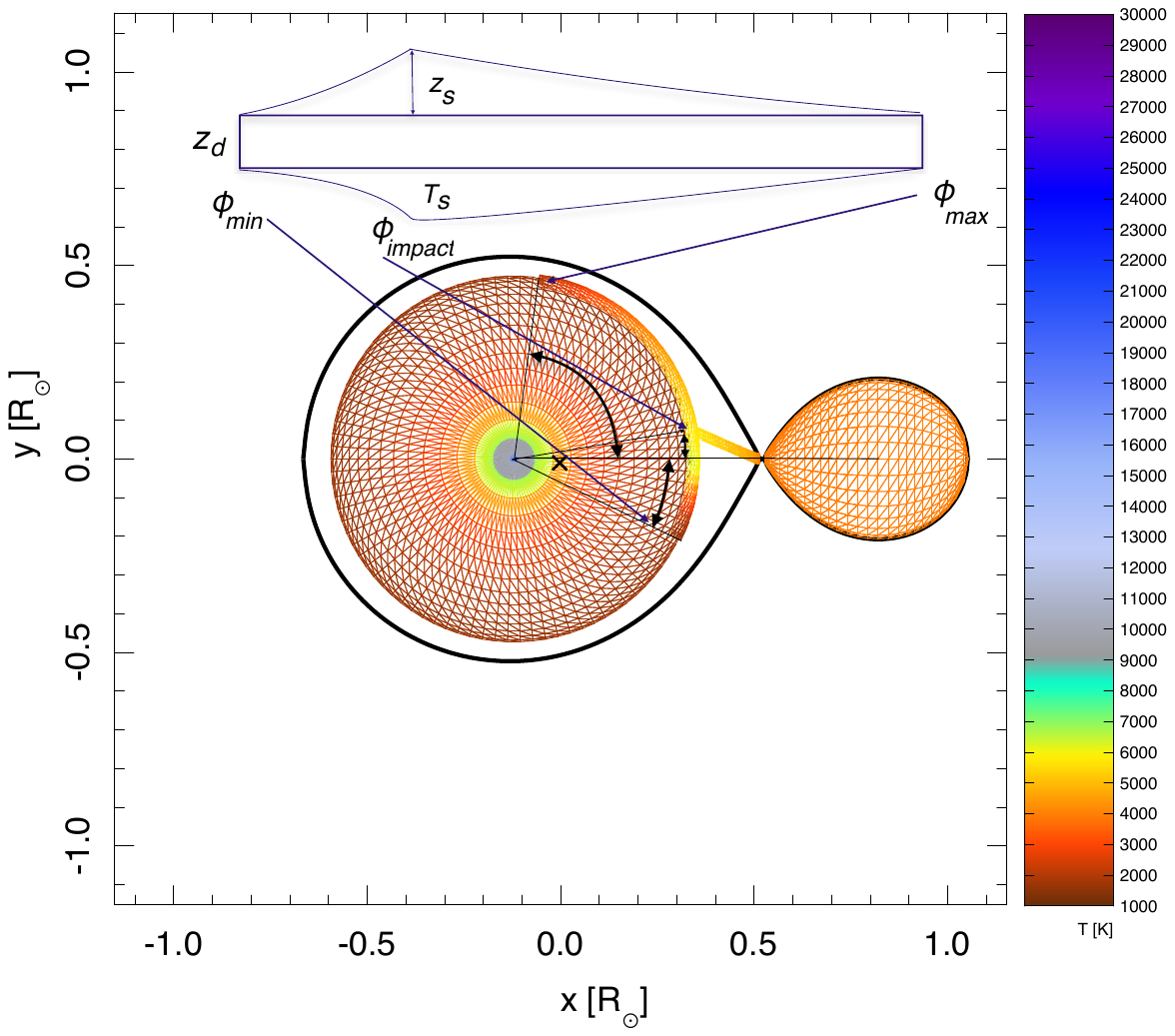}
  \caption{Model of CzeV404 Her including the definition of the hot-spot parameters (see \citep{2021A&A...652A..49K}).}
  \label{figSpot}
\end{figure}

\subsection{The hot spots/lines}

We used a complex model for the spot, as presented in Fig.~\ref{figSpot}. The impact of the stream with the outer
edge of the disk forms a shock-heated trailing arc along the rim. The vertical size $z_{s}$ and the temperature $T_{s}$ of the arc-like hot spot are not uniform; it is wider and hotter at the impact point ($\phi_{impact}$), quickly increases before the impact point, and slowly
declines after one in the temperature and the vertical extension,
according to an arbitrarily selected manner described by the following equations:
\begin{eqnarray*}
T_{s, i} & = & T_d(1.0+ \gamma_{T,i}f(\phi)^{\alpha_{T,i}}) \\
z_{s, i} & = & z_d(1.0+ \gamma_{z,i}f(\phi)^{\alpha_{z,i}})
\end{eqnarray*}
where $\varphi$ is the angle between the line connecting the stars in
conjunction and the direction of the spot viewed from the centre
of mass; $f(\varphi) = \hat{a}\varphi+\hat{b}$  is a linear function with constants $\hat{a}$ and $\hat{b}$ defined by a 
boundary conditions of  $f(\varphi_{min}) =0$, $f(\varphi_{impact})= 1$
 and $f(\varphi_{max}) =0$, and $\gamma_T$ , $\gamma_z$ are free parameters. The index
$i= 1,2$ corresponds to parts of the hot spot before and after the
impact point.

A simple modification of the code allows the inclusion of any number of spots, both at the outer and inner radii of the disk. The boundary layer (the region in which gas moving at Keplerian velocity in the disk is decelerated to match the surface velocity of the primary) can be included as a contribution from an elongated hot spot at the inner radius of the disk, provided that it extends to the surface of the primary.

\subsubsection{The spiral arms in the accretion disk}
The two humps per orbital period in the light curve of dwarf novae with small mass ratio 
was predicted in \citep{1979MNRAS.186..799L} as a result of formation of spiral density
waves in the outskirts of the disk under the influence of 2:1 resonance. The code includes this option
following \citep{2004ApJ...606L.139H}. The spiral pattern is described by multiplying thickness of the disk h$_\mathrm{disk}$ with $z_h$, as defined below:
\begin{equation}
z_h   = max\left(1, \frac{\xi}{(\sqrt{(r/R_{\mathrm{disk,out}})-exp(-\eta(\phi - \delta - i_{sp})))^2+\epsilon^2}}\right),
\end{equation}
where $\eta$ is the inverse of pitch angle, $\delta$ - position angle of the spirals against the binary components, $\epsilon$ is the width of the spiral pattern, $i_{sp}$ = 0 degree for the first spiral and 180 degree for 
the second, $r$ and $\phi$ the disk element coordinates in the polar system.
In addition to the original model we put an excess of the effective temperature $\gamma_{spiral}$ in the spiral  ( $z_h > 1$) defined as 
\begin{equation}
T(r,\phi) = T_{d}  ( 1+ \gamma_{spiral} \times  z_h(r, \phi)), 
\end{equation}

\subsection{Magnetic field and spot at the primary}

The code allows the definition of a spot on the primary surface and a dipole magnetic field. 
The spot is characterized by its radius (expressed as a fraction of the primary radius) and by two angular parameters: the inclination angle between the rotational axis and the spot centre, and the azimuthal angle. 
The initial azimuth is defined with respect to the direction of the terminal point of the accretion stream. 
If a magnetic field is present, it is assumed to originate from the centre of the spot and to share the same inclination. 
The radius of the magnetic field corresponds to the inner radius of the accretion disk. 
When the primary rotates with its intrinsic rotational period, the magnetic field rotates accordingly.  

It is possible to trace the trajectories of accreting material within the magnetosphere along the magnetic field lines. 
However, the computation of the radiative flux from such magnetically guided accretion streams represents a complex problem and is not yet implemented in the current version of the code.

\section{Code structure and running}

The code was written in the C programming language and employs the {\it plplot}\footnote{\url{https://plplot.sourceforge.net}} library for system visualization and graphical output. 
It consists of a main program and a set of modular subroutines, each responsible for the computation and visualization of a specific system component.  

The current version of the code includes the following functional modules:  
\begin{itemize}
    \item[a)] {\bf Main body:} {\tt CVlab202}
    \item[b)] {\bf System components:} {\tt primary}, {\tt secondary}, {\tt stream}, {\tt disk}, {\tt streamMF}, {\tt primaryMF}
    \item[c)] {\bf Radiation components:} {\tt flux}, {\tt blackbody}, {\tt filter}, {\tt limbdarking}, {\tt limbdarkingWD}
    \item[d)] {\bf Geometry components:} {\tt direction}, {\tt line}, {\tt triangl}, {\tt geometry}, {\tt minmax}
    \item[e)] {\bf Input/output components:} {\tt input}, {\tt systempar}, {\tt printSys}, {\tt SpType}
    \item[f)] {\bf Visualization components:} {\tt view0}, {\tt view1}, {\tt view2}
\end{itemize}

\begin{figure}
\includegraphics[width=1.0\textwidth]{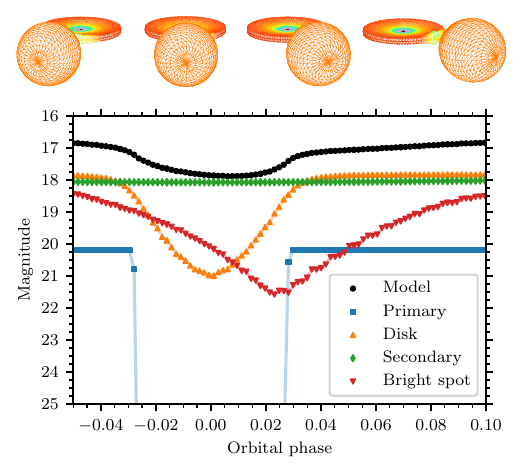}
  \caption{The contribution of different components to the total flux.
The top side of the plot presents an artistic view of the system in
corresponding orbital phases. The lack of continuity in the bright
spot curve is due to a pixelation of our model. (from \citep{2023ApJ...950...47K})}
  \label{figEclipse}
\end{figure}

\subsection{Code input parameters}

The input parameters of the code are specified in the {\it param.ini} file, which currently includes 55 parameters. 
These parameters can be grouped into several categories: 
(1) parameters defining the calculation sequence; 
(2) parameters specifying the optical and infrared bands for light-curve modeling; 
(3) the distance to the object and the interstellar absorption; 
(4) general system parameters such as the mass of the primary, the system inclination, and the mass ratio; 
(5) parameters describing the accretion disk, the number and properties of spots, and the presence of spiral arms; 
(6) parameters defining the type and rotation of the primary, the magnetic field, and the possible presence of a surface spot; 
and (7) technical parameters controlling graphics and output options.  

The code can be executed as a stand-alone program or embedded within a script or a subroutine of another computational package.

\begin{figure}[t]
\includegraphics[width=1.0\textwidth]{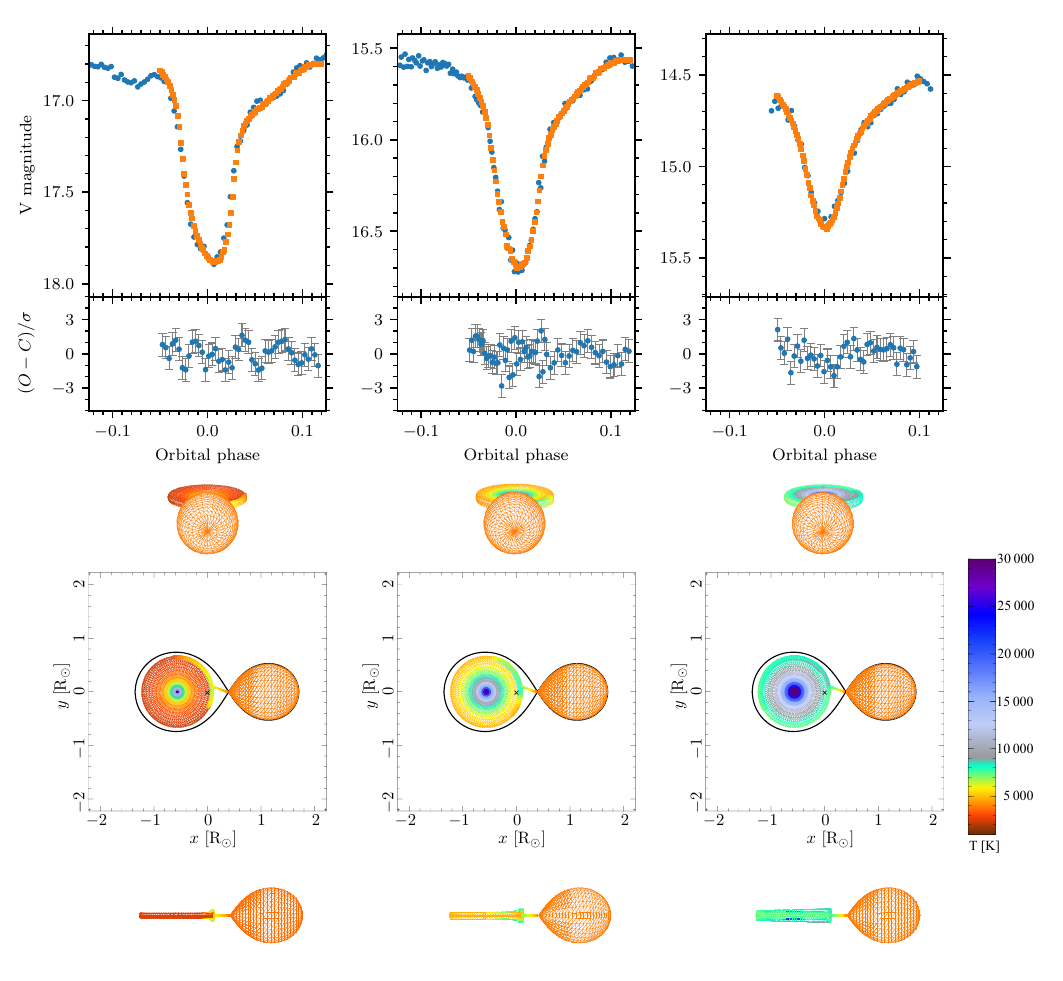}
  \caption{{\it Upper part}: Results of the light curve fit of AY~Psc eclipses in quiescence/low ({\it left}), standstill  ({\it middle})
and in outburst/high ({\it right}) state. 
{\it Bottom}: Models of AY~Psc in quiescence ({\it left}), standstill ({\it middle}) and in outburst ({\it right}).  The colors mark the effective temperature in Kelvin of the system elements ({\it right bar}). The upper panels correspond to orbital phase $\varphi=0$, the middle ones are views on the geometry of the system at the orbital phase of $\varphi=0.25$ from 
the inclination angle $i=0.0$, and the lower panels correspond o $\varphi=0.25$  and inclination angle $i=90.0$ degree.   (from \citep{2023ApJ...950...47K})}
  \label{figEclipse2}
\end{figure}

\subsection{The code output}

Depending on the input parameters, the code produces the following output:  
a file {\it system.out}, which contains general information about the performed calculations;  
a file {\it Phot.dat}, which includes the resulting model light curve (see Fig.~\ref{figEclipse});  
a set of files describing the 3D geometry of each system component as visible to the observer;  
and several PDF files containing graphical visualizations of the system and its components.  
At minimum, the code can compute the total flux and generate a visual representation of the system at a specified orbital phase, as seen by the observer. The typical computation time for a single orbital phase of a system that includes all possible components is $\lesssim 2$~minutes.  
By using scripting and multicore computing resources, it is possible to generate complete model light curves within a reasonable total computation time.

\subsection{The fitting of observational data}

The fitting of the observational data is possible by using an additional minimization 
program (for an example exploring the gradient descent method \citep{Press2007})  which minimizes
$$\chi^2 = \sum_i^n\left(\frac{mag_{mod,i}-mag_{obs,i}}{\Delta mag_{obs,i} }\right)^2, $$
the $mag_{obs,i}$, $mag_{mod,i}$ is observational and model data in corresponding
band and phases ($i$). The minimization can be applied to any user-selected subset of model parameters.

\section{Examples of code applications}

There are a number publications with application of different versions of the code to analyses of  binary systems as nova-like\citep{2017MNRAS.470.1960H, 2020MNRAS.497.1475S, 2023MNRAS.521.5846M} and dwarf novae cataclysmic variables\citep{2013A&A...549A..77Z, 2013MNRAS.432.2596N, 2015AcPPP...2...41Z, 2021A&A...652A..49K, 2021ApJ...918...58A, 2023ApJ...950...47K, 2023MNRAS.526.5110E, 2025A&A...699A..81K},  millisecond pulsars\citep{2019MNRAS.489.5547Z, 2024MNRAS.527.6712Z, 2024A&A...690A.173B, 2025A&A...693A.158K, 2025A&A...702A.256B}, a low-mass X-ray binary \citep{2014MNRAS.445.2424N}, and a B[e] star \citep{2024ApJ...968...52N}.  I give an example of results obtained for Z-Cam type system AY Psc obtained recently by us \citep{2023ApJ...950...47K}
in Fig.~\ref{figEclipse2}. The code provided a good agreement with another modelling tool also (see \citep{2023MNRAS.526.5110E}). 

\section{Limitations of the code}

Firstly, the code is based on an idealized geometric representation of the system, as described above.  
Naturally, the shapes of the accretion disk and the hot spot adopted in the model are strong simplifications of the real physical structures.  
Secondly, the number of surface elements used to describe each component is limited by the available computer memory and computation time. 
The code can fit only binary system with circular orbits and a secondary with rotation  locked to the orbital period. 

Another significant limitation arises from the assumption that each emitting element radiates as a black body.  
This approximation complicates the fitting of multi-band observational data, since the intrinsic spectra of the components deviate from the black-body distribution.  
Consequently, in optically thin regions, the derived effective temperature represents only the radiation intensity and does not correspond to the kinetic temperature of the emitting plasma.  
Therefore, differences in the effective temperature derived from various photometric bands indicate spectral deviations from the black-body law.

In general, for this type of modeling, it is essential to clearly recognize that the interpretation of the obtained results is constrained by the underlying assumptions of the model.

\section{Summary and Future Development}

CVlab V2.0 is designed to fit the light curves of close binary systems such as cataclysmic variables (CVs), AM CVn stars, Algol-type binaries, millisecond pulsars, and low-mass X-ray binaries (LMXBs). In certain cases, it can also be applied to model the observational data of ordinary binary stars with ZAMS or evolved components, provided that the inherent limitations of the code are taken into account.
The primary purpose of the code is to determine the fundamental characteristics of binary systems, including the component's masses, radii, effective temperatures, and the parameters of accretion disks together with any structures present within them. In addition, the code provides visualization of the system geometry, the distribution of effective temperature across the components, and the velocity field within the system.
Knowledge of the velocity field within the system will also enable the construction of synthetic Doppler maps of the accretion structures. 

Future development of the code aims to incorporate a more realistic accretion-disk structure computed using hydrodynamic simulations, as well as stellar and disk-atmosphere models to generate more realistic  system emission.

The code is not yet publicly available, as it is still undergoing development and testing. However, it can be used to analyze specific data upon direct request to the developer (zhar@astro.unam.mx).

\acknowledgments
The author acknowledges DGAPA-PAPIIT grant  IN105826














\bibliographystyle{JHEP}
\bibliography{ref}

\end{document}